\documentclass[12pt]{article}
\usepackage{authblk} 
\usepackage{geometry}
\usepackage{graphicx} 
\usepackage{xcolor}   
\usepackage{mathrsfs} 
\usepackage{tabularx} 
\usepackage{caption}  
\usepackage{subcaption} 
\usepackage{amsmath}  
\usepackage[noadjust]{cite} 
\usepackage{placeins} 
\usepackage{float}    
\usepackage{hyperref} 
\usepackage{cleveref} 
\usepackage{placeins}

\title{Finite and Asymptotic Key Analysis for CubeSat-Based BB84 QKD with Elliptical Beam Approximation}
\author[1]{Muskan\thanks{Corresponding Author E-mail: muskan.1@iitj.ac.in}}
\author[1,2]{Arindam Dutta\thanks{E-mail: arindamsalt@gmail.com}}
\author[1]{Subhashish Banerjee\thanks{E-mail: subhashish@iitj.ac.in}}

\affil[1]{Department of Physics, Indian Institute of Technology, Jodhpur, Jodhpur 342030, Rajasthan, India}
\affil[2]{Department of Physics and Materials Science and Engineering, Jaypee Institute of Information Technology, A 10, Sector 62, Noida, UP-201309, India}

\begin{document}

\maketitle

\begin{abstract}
 {Quantum Key Distribution (QKD) via CubeSats presents a promising pathway for secure global communication. However, realistic modeling of the satellite-to-ground optical link remains a challenge due to beam divergence, atmospheric turbulence and finite key effects. In this work, we employed an elliptical beam approximation model for the downlink channel to more accurately characterize transmittance variations compared to conventional circular beam models. This refined approach facilitates a precise estimation of finite and asymptotic key rates under realistic weather conditions, for weak coherent pulse (WCP)-based efficient BB84 and standard BB84 protocols using two decoy states in a CubeSat-based downlink scenario. Our results reveal that the efficient BB84 protocol consistently yields higher key rates, particularly under turbulent conditions, making it more suitable for practical deployment. Additionally, we analyze  probability distribution of key rate (PDR) across different zenith angles demonstrating the robustness of elliptical beam modeling for CubeSat-based QKD. This study provides an improved framework for designing realistic CubeSat-based QKD systems, bridging theoretical models and experimental feasibility.}

\end{abstract}
\providecommand{\keywords}[1]
{
  \small  
  \textbf{\textit{Keywords---}} #1
}

\section{Introduction}

Quantum Key Distribution (QKD) and quantum communication provide a framework for secure information transmission, with quantum identity authentication ensuring the verification of legitimate users \cite{CS01, DP22, DPA22}.  Photons serve as fundamental carriers of quantum information, facilitating high communication rates in optical fiber networks. However, extending quantum communication over long distances remains challenging due to inevitable photon loss during propagation \cite{MPR+19}. While quantum repeaters have been proposed as a potential solution \cite{ATL15, ZPD+18, SGL18}, significant technological hurdles remain. In contrast, satellite-based free-space links currently offer the most viable approach for long-distance QKD ~\cite{ P3021}. These systems leverage advancements in satellite and optical communication technologies, with various techniques developed to mitigate the effects of noisy quantum channels \cite{SB18}. Extensive feasibility studies ~\cite{ELH+21, CSH+22} and experimental demonstrations indicate that satellite-based QKD is technologically mature and ready for deployment.  {Asymptotic key rate analysis plays a crucial role in optimizing the performance of such satellite-based QKD systems, as demonstrated in \cite{ABN21}, which focuses on optimizing  signal and decoy intensities to maximize secure key generation. Furthermore, the asymptotic key rate analysis for both continuous-variable and discrete-variable QKD in the presence of bypass channels is presented in \cite{GBL+23}, deriving key rate bounds under restricted eavesdropping scenarios and demonstrating that limiting Eve’s access can enhance the asymptotic key rate, particularly in high-loss conditions.}\\ 
 {
As satellite-based QKD becomes increasingly viable, it becomes crucial to accurately model the security of key generation under realistic constraints, particularly in finite-key scenarios. Recent advancements in finite-key analysis for QKD have led to significant improvements in secure key rate estimation by addressing statistical fluctuations and optimizing security parameters \cite{PKH23}. Initially, numerical methods were developed to compute secure key rates by incorporating statistical uncertainties in key generation \cite{BGK+20}. Subsequently, semidefinite programming (SDP) was introduced to establish a generalized finite-key framework applicable to various QKD protocols, including device-independent schemes \cite{GLL21}. Further research extended these analyses to satellite-based QKD, evaluating finite-key effects in the presence of orbital dynamics, channel losses, and statistical corrections \cite{SBM+22,ELJ+21,DTR+21}. More recently, finite-key analysis in satellite QKD has incorporated real-world constraints, including hardware limitations, link efficiency, and environmental factors \cite{SBM+23}.}\\
 {
Building upon recent advantages this work presents the performance of tight statistical techniques for parameter estimation and error correction to compute the finite-key block size as proposed in \cite{LCW+14, SBM+22} and the asymptotic key rate for weak coherent pulse (WCP)-based efficient BB84 and standard BB84 protocols using the elliptical beam model in a CubeSat-based QKD downlink scenario. The focus of space-based quantum communication has increasingly shifted towards smaller satellites, particularly CubeSats, which are the most prevalent type of nanosatellite. Recent CubeSat missions, such as Jinan-1 \cite{LCJ+24}, have demonstrated the feasibility of secure quantum communication using compact satellite platforms. We have considered a CubeSat in Low Earth Orbit (LEO) at a constant altitude of 400 km, as CubeSats in LEO are well-suited for QKD due to their cost-effectiveness \cite{ZSH+24}, lower channel loss, and practical feasibility for secure quantum communication. Compared to Medium Earth Orbit (MEO) and Geostationary Orbit (GEO) satellites, the shorter distances in LEO reduce free-space losses and enhance photon detection probabilities, thereby improving key rate generation \cite{LVM+23, SLM+22, MTP+24}. To accurately model beam propagation and its impact on secure key rates, this study employs the elliptical beam approximation \cite{LKB19}, which offers a more realistic representation of beam dynamics in CubeSat-based QKD links compared to the circular beam model. Unlike prior studies on CubeSat-based QKD \cite{ZSH+24}, which primarily account for diffraction losses and background noise while neglecting detailed turbulence modeling, our approach explicitly incorporates beam wandering, elliptical deformations of the beam profile, and extinction losses from backscattering and absorption. This enhanced turbulence model enables precise transmittance evaluation through a receiving aperture by considering parameters such as centroid position, semi-axes of the elliptical profile, and orientation angle. This model is particularly useful for assessing beam fluctuations under different weather conditions, where turbulence and scattering particles, such as haze and fog, affect propagation \cite{VSV16} . Furthermore, while prior works such as \cite{SLM+22} focus on the implementation of an optical ground station and CubeSat-based entanglement distribution, they do not explore the implications of finite key rate analysis on security, which are crucial for practical QKD deployments. In contrast, our study specifically addresses these finite key rate analyses, providing a more comprehensive assessment of security in CubeSat-based QKD scenarios.}\\
 {
 We have opted for a downlink configuration over an uplink due to its lower transmission losses \cite{LKB19}, making it more suitable for CubeSat applications, particularly for the BB84 protocol \cite{GN22}, which is one of the most widely implemented QKD schemes due to its simplicity and proven security \cite{SP00}. The key rate analysis is conducted for both efficient and standard BB84 protocols employing elliptical beam approximation model from satellite-to-ground station link under various weather conditions. Additionally, the probability distributions of key rates (PDR) are examined with respect to zenith angles to provide a comprehensive understanding of CubeSat-based QKD performance. The BB84 protocol is opted for simulation due to its well-established security, ease of implementation, and extensive experimental validation \cite{SP00,OKM+24}. While more advanced protocols exist, BB84 remains widely adopted for its resilience against eavesdropping attacks and compatibility with various quantum communication architectures \cite{LSL22}. Recently, the BB84 protocol has been implemented for CubeSat-based QKD in \cite{ZSH+24}, demonstrating its effectiveness and suitability for such applications. Owing to its widespread adoption for security, ease of implementation, and experimental validation, we employ the two decoy-state BB84 protocol in our work for finite and asymptotic key analyses using elliptical beam model in the CubeSat-based scenario.
\\
To further enhance key generation efficiency while maintaining security, this study incorporates both efficient and standard BB84 protocols using two decoy states. The efficient BB84 protocol employs a biased basis choice, improving the sifting ratio and increasing the fraction of signals used for key generation \cite{SBM+22}. This is particularly advantageous in CubeSat, where limited transmission time and high channel loss due to beam divergence, atmospheric attenuation, and background noise impact key rates. By optimizing basis selection, efficient BB84 enhances raw key generation while maintaining security, making it well-suited for CubeSat-based QKD, as demonstrated in \cite{MTP+24}, which analyzes a CubeSat in a 500 km sun-synchronous orbit (SSO) performing downlink QKD to an optical ground. In our study, while efficient BB84 maximizes key rates, standard BB84 remains relevant due to its symmetric basis selection, rigorous security framework, and broad applicability across quantum communication. Furthermore, \cite{MTP+24} proposes a payload for downlink QKD but is limited to a simplified version of BB84 using only one decoy state. In contrast, our study incorporates both efficient and standard BB84 protocols using two decoy states, enhancing security and robustness against eavesdropping attacks. While atmospheric losses are considered in their work, there is no quantitative study on how different atmospheric conditions impact key rates, a key aspect addressed in our analysis. Moreover, our work examines the PDR under varying atmospheric conditions, adding a statistical depth absent in their study. By analyzing both protocols and incorporating finite and asymptotic key rate analyses, elliptical beam propagation modeling, assessments under various atmospheric conditions, and PDR evaluation across different zenith angles, our study provides a more comprehensive and realistic assessment of CubeSat-based QKD performance. }\\

The remainder of this paper is organized as follows: Section 2 presents an explanation of the decoy-state-based efficient BB84 and standard BB84 protocols. Additionally, it explores the finite key rate analysis and the asymptotic key rate analysis for these protocols. Moreover, we explore the impact of atmospheric conditions on CubeSat communication links and analyze the elliptical beam deformation approximation at the receiver. Section 3 provides a detailed assessment of the performance of the efficient and standard BB84 protocols, supplemented with illustrative results obtained from simulations. Finally, Section 4 concludes the paper by summarizing the key findings and discussing their implications. The Appendix includes additional details that support the analyses presented in the main sections.

\section{Preliminaries}
\label{Preliminaries}
\subsection {Protocol Description: efficient BB84 and standard BB84 (using two decoy settings)}
The BB84 QKD protocol \cite{BB14} has gained extensive adoption due to its straightforward design, robust performance, and theoretically sound security guarantees. Despite this, practical deployments of BB84 typically deviate from the idealized single-photon sources; instead, weak coherent laser pulses are favored for their widespread availability and implementation feasibility. Although such laser sources enhance repetition rates compared to current single-photon emitters, they also render BB84 susceptible to PNS attacks that exploit the multiphoton components of emitted pulses. Decoy state protocols effectively mitigate PNS vulnerabilities and increase resilience to substantial channel losses, requiring only minimal alterations to  BB84 implementations. To mitigate the challenges associated with multiphoton components and channel losses, we utilize both standard and efficient BB84 protocols augmented with decoy-state settings. The standard BB84 protocol with decoy states provides a robust framework for secure key generation, while the efficient version improves key rates by maximizing the utilization of transmitted quantum states. In the following subsections, we systematically analyze the performance of these protocols under practical conditions.\\

\subsubsection{ Efficient BB84 and standard BB84 protocol (using two decoy settings)}  { In the efficient BB84 protocol \cite{LCA05}, Alice and Bob select between the Z basis (\(|0\rangle, |1\rangle\)) and the X basis (\(|+\rangle, |-\rangle\)) with asymmetric probabilities \( c_X \) and \( 1 - c_X \). The Z basis is used for parameter estimation, while the X basis is used for key generation. The protocol employs phase-randomized laser pulses and a decoy-state method with three intensity levels \( \mu_1, \mu_2, \mu_3 \) satisfying \( \mu_1 > \mu_2 + \mu_3 \) and \( \mu_2 > \mu_3 \geq 0 \) \cite{LCW+14}.}\\  

 {\textit{Preparation and measurement}}:   {Alice randomly selects a bit \( m_i \), a basis \( A_i \in \{X, Z\} \) with probabilities \( c_X, 1 - c_X \), and an intensity \( q_i \in \mathcal{Q} = \{\mu_1, \mu_2, \mu_3\} \) with probabilities \( p_{\mu_1}, p_{\mu_2}, p_{\mu_3} \). She then transmits a weak laser pulse to Bob, who selects a basis \( B_i \) with the same probabilities and records the measurement result \( m_i' \). Bob’s outcomes include \( \{0, 1, \emptyset, \bot\} \), where \( \emptyset \) indicates no detection, and \( \bot \) (double detection) results in a random bit assignment.} \\ 

 {\textit{Basis reconciliation and raw key generation}}:   {Alice and Bob publicly announce bases and intensities, defining sets \( \chi_q, \mathscr{Z}_q = \{ i : A_i = B_i \in \{X, Z\}, q_i = q, m_i' \neq \emptyset \} \). If \( |\chi_q| \geq n_{X,q} \) and \( |\mathscr{Z}_q| \geq n_{Z,q} \) for all \( q \), they generate a raw key pair \( (X_{\text{Alice}}, X_{\text{Bob}}) \) by sampling \( w_X = \sum_{q \in \mathcal{Q}} w_{X,q} \) from \( \bigcup_{q \in \mathcal{Q}} \chi_q \).} \\ 

 {\textit{Error estimation and post-processing}}:  
 {Alice and Bob estimate errors using \( \mathscr{Z}_q \), determining bit errors \( m_{Z,q} \), vacuum (\( y_{X,0} \)) events,  single-photon (\( y_{X,1} \)) events and phase errors \( f_{X,1} \). If the phase error rate \( \varphi_X = f_{X,1}/y_{X,1} \) exceeds the threshold \( \varphi_{\text{tol}} \), they abort. Otherwise, they proceed with error correction (leaking at most \( \lambda_{\text{ec}} \) bits), error verification (leaking \( \log_2(1/\varepsilon_{\text{hash}}) \) bits), and privacy amplification to extract a final secret key \( (S_A, S_B) \) of length \( l \).}  \\
 {In the standard BB84 protocol with two decoy states \cite{LMC05}, Alice and Bob choose bases randomly, uniformly, and independently. Unlike the efficient BB84 protocol, it employs an unbiased basis choice and uses both bases for key generation and parameter estimation \cite{SBM+22}.}\\
 
The calculation cost of the efficient BB84 and standard BB84 protocols differs due to the asymmetric basis selection in the efficient variant, which optimizes key generation and error estimation, making it more efficient than standard BB84 \cite{LCA05}. Efficient BB84 reduces data loss through biased basis selection, lowering sifting and error correction costs, while the information leakage, dependent on the data block size \cite{SBM+22}, slightly increases due to a higher raw key rate. Additionally, its phase error estimation is simpler, requiring fewer computational resources than standard BB84. Table \ref{table1} presents a comparative analysis of the calculation costs for both protocols, outlining the key differences across various computational steps.\\
\begin{table}[H]
    \centering
     \caption{Calculation cost comparison of Efficient BB84 and Standard BB84 protocols.}
    \begin{tabularx}{\textwidth}{|X|X|X|}
        \hline
        \textbf{Computation Step} & \textbf{Efficient BB84} & \textbf{Standard BB84} \\
        \hline
        \textbf{Basis Reconciliation}  & The efficiency of this scheme is asymptotically twice that of the standard BB84 protocol due to its biased basis selection, reducing the computational cost associated with sifting by minimizing data rejection. & Approximately 50\% of the data is discarded due to Bob's random basis selection, which results in incorrect measurements half of the time.
 \\
        \hline
        \textbf{Error Estimation} & In our analysis, the QBER for the efficient BB84 protocol is determined to be approximately 5\%, requiring a lower error correction cost. & Likewise, the QBER for the standard BB84 protocol is observed to be around 10\%,  leading to a higher computational cost for error correction. \\   
        \hline
      \textbf{Error Correction} & In the finite key regime, the information leakage during error correction is approximately 0.0004375 bits per pulse, depending on the data block size. Higher raw key rate in efficient BB84 leads to increased leakage. & Here, the leakage is around 0.00034 bits per pulse, as more data is discarded during sifting, resulting in a lower raw key rate and reduced leakage. \\  
        \hline
        \textbf{Phase Error Estimation} & Involves calculating phase error rate only from the X basis, reducing computational complexity. & Requires phase error estimation from both X and Z bases, increasing computational burden. \\
        \hline
    \end{tabularx}
    \label{table1}
\end{table}

\subsection{Finite key rate and asymptotic analysis} 
\label{Finite key rate and Asymptotic analysis}
This work focuses on examining the finite key rate and asymptotic key rate for both efficient and standard BB84 protocols for two decoy states in CubeSats. CubeSats, with their ability to operate in low-Earth orbit, play a crucial role in enabling QKD by providing a practical platform for secure communication. Their compact design and suitability for downlink scenarios make them an ideal platform for analyzing the performance of key rates under practical constraints. Detailed insights into this analysis are provided in the subsequent subsections.

\subsubsection{Finite key rate analysis for decoy-state BB84 in CubeSat systems}  
 {The security of decoy-state QKD was initially developed under the assumption of the asymptotic-key regime \cite{W05, LMC05}. However, for practical implementations with finite data sizes, uncertainties in the channel parameters must be taken into account \cite{MQZ+05, HHH+07, CS09}. Early methods addressing finite-key effects relied on Gaussian approximations to quantify the discrepancy between asymptotic and finite-key results \cite{ZZR+17}. These approaches, however, limited the security analysis to collective and coherent attacks.  Later advancements extended the finite-key security analysis to include more general attack strategies \cite{HN14}, utilizing bounds such as the multiplicative Chernoff bound \cite{CXC+14, ZZR+17} and Hoeffding’s Inequality \cite{LCW+14} to quantify statistical fluctuations. A composable finite-key analysis for decoy-state efficient BB84 using the multiplicative Chernoff bound was introduced in \cite{YZG+20}, providing tighter security bounds and improving the estimation of key parameters.}\\
 {In this work, we extend these finite-key rate analyses to CubeSat-based QKD}, where CubeSats in LEO enable secure key exchange over free-space optical links. The overpass duration of a CubeSat, or time window, determines the total number of pulses transmitted during a single pass. Let \( N_{p} \) be the total number of pulses transmitted during a CubeSat pass, which depends on the source repetition rate and the duration of the CubeSat overpass, referred to as the time window. For a maximum zenith angle of \( 80^\circ \), the time window is typically limited to approximately \( 440~\mathrm{seconds} \) \cite{SBM+22}. The finite key rate
 {measured in bits per pulse} for a single pass, based on the efficient BB84 protocol, is then determined as \cite{SBM+22} \\
\begin{equation}
 R^{eff} = \frac{l}{N_p} =\left\lfloor y_{X,0}+y_{X,1}(1-h(\varphi_X))-\lambda_{ec}-6\log_2\frac{21}{\varepsilon_{sec}}-\log_2\frac{2}{\varepsilon_{corr}}\right\rfloor.
 \label{eq_reff}
\end{equation}
Here $y_{X,0}$, $y_{X,1}$, and $\varphi_X$ represent the vacuum yield, single-photon yield, and phase error rate in the $X$-basis, respectively  { and \( h(x) := -x \log_2 x - (1 - x) \log_2 (1 - x) \) is the binary entropy function.} The amount of information leakage is quantified by $\lambda_{ec}$, which is considered during privacy amplification. In the finite key regime, this leakage is fundamentally bounded by $\lambda_{\text{ec}} \leq \log|\mathcal{W}|$, where $\mathcal{W}$ denotes the set of syndromes involved in the information reconciliation process. We utilized an estimate of $\lambda_{ec}$ that varies with the block size, as described  below  \cite{TMP+17}
\begin{equation}
\begin{split}
\lambda_{ec} = & \, n_X h(Q) + n_X (1-Q) \log\left[\frac{(1-Q)}{Q}\right] \\
& - \left(F^{-1}(\varepsilon_{corr}; n_X, 1-Q) - 1\right) \log\left[\frac{(1-Q)}{Q}\right] \\
& - \frac{1}{2} \log(n_X) - \log\Big(\frac{1}{\varepsilon_{corr}}\Big).
\end{split}
\label{eq_lambda_ec}
\end{equation}
Here, $n_X$ represents the data block size, $Q$ denotes the quantum bit error rate (QBER) \cite{SBM+21}, and $F^{-1}$ refers to the inverse of the cumulative distribution function of the Binomial distribution. This definition is used to evaluate the quantity of information that is leaked during the error correction process in the finite key regime. The protocol's reliability and security are characterized by two parameters, $\varepsilon_{corr}$ and $\varepsilon_{sec}$. A protocol is considered $\varepsilon$-secure if it satisfies $\varepsilon = \varepsilon_{corr} +\varepsilon_{sec}$, where it is $\varepsilon_{corr}$-correct and $\varepsilon_{sec}$-secret.

We now analyze the standard BB84 protocol using WCPs with two-decoy-states. The finite key rate for single pass for standard BB84 protocol can be expressed as follows \cite{SBM+22}: 
\begin{equation}
\begin{split} 
R^s = \frac{l}{N_p} = \frac{1}{2} \left\lfloor y_{X,0} + y_{X,1} \left( 1 - h(\varphi_X) \right) - \lambda_{ecX} + y_{Z,0} + y_{Z,1} \left( 1 - h(\varphi_Z) \right) \right. \\
\left. - \lambda_{ecZ} - (12\log_2 \frac{21}{\varepsilon_{sec}} - 2\log_2 \frac{2}{\varepsilon_{corr}}\Big) \right\rfloor.
\end{split}
\label{eq_rs}
\end{equation}

Satellite-based quantum communication systems are significantly impacted by finite statistical effects due to the limited duration of transmission windows. We use improved analysis of \cite{SBM+22} in modelling statistical fluctuations arising from finite statistics. This enhances the robustness of the secret key rate and incorporates a finite-statistics correction term, denoted as \( \mathrm{\delta}^{\pm}_{X(Z),j} \). This correction term is defined using the inverse multiplicative Chernoff bound \cite{YZG+20, ZZR+17}. Specifically let Y denotes a sum of $\mathcal{T}$ independent Bernoulli samples, which need not be identical. Denote $y^\infty$ as the expectation value of Y, with $y$ the observed value for Y. The extent of the discrepancy between the observed and expected values is influenced by the available statistics. To quantify this deviation, the probability that \( y \leq y^\infty + \delta_Y^{+} \) is less than a fixed positive constant \( \epsilon > 0 \), and the probability that \( y \geq y^\infty - \delta_Y^{-} \) is less than \( \epsilon \). This is achieved through setting
\begin{equation}
    \mathrm{\delta}_Y^{+} = \beta + \sqrt{2\beta y + \beta^2}, \quad \mathrm{\delta}_Y^{-} = \frac{\beta}{2} + \sqrt{2\beta y + \frac{\beta^2}{4}},
    \label{eq_delta}
\end{equation}
where $\beta=\ln{\frac{1}{\epsilon}}$ \cite{YZG+20}. Hence, we define the following finite sample size data block size \cite{SBM+22, SBM+21}.
\begin{equation}
\begin{split}
    n^{\pm}_{X(Z),j} = \frac{e^j}{p_j}\Big[  n_{X(Z),j}\pm \mathrm{\delta}^{\pm}_{n_{X(Z),j}}\Big] ,\\
    m^{\pm}_{X(Z),j} = \frac{e^j}{p_j}\Big[  m_{X(Z),j}\pm \mathrm{\delta}^{\pm}_{m_{X(Z),j}}\Big],
\end{split}    
\label{eq_nene}
\end{equation}
for the number of events and errors respectively in the $X(Z)$ basis. From this the vaccum and single photon yields, and the phase error rate of single photon events are defined as given in \cite{LCW+14}. The number of vacuum events in $X_A$ satisfies
\begin{equation}
    y_{X,0}\geq \frac{\tau_0\mu_2 n^-_{X,\mu_3}-\mu_3 n^+_{X,\mu_2}}{\mu_2-\mu_3},
    \label{eq_yx0}
\end{equation}
where $\tau_n:=\sum_{j \in \kappa}e^{-j}j^np_j/n!$ is the probability that Alice sends n-photon state. The number of single photon events in $X_A$ is 
\begin{equation}
   y_{X,1} \geq \frac{\tau_1 \mu_1 \left[n_{X,\mu_2}^- - n_{X,\mu_3}^+ - \frac{\mu_2^2 - \mu_3^2}{\mu_1^2} \left(n_{X,\mu_1}^+ - \frac{y_{X,0}}{\tau_0}\right)\right]}{\mu_1 (\mu_2 - \mu_3) - (\mu_2^2 - \mu_3^2)}.
   \label{eq_yx1}
\end{equation}
 The number of vacuum events, \(y_{Z,0}\), and the number of single-photon events, \(y_{Z,1}\), using Eqs. (\ref{eq_yx0}) and (\ref{eq_yx1}) can also be defined. Additionally, the number of bit errors, \(v_{Z,1}\), associated with the single-photon events in the \(Z\)-basis is also computed. It is given by
\begin{equation}
    v_{Z,1}\leq \tau_1 \frac{m_{Z,\mu_2}^+-m_{Z,\mu_3}^-}{\mu_2-\mu_3}. 
    \label{eq_vz1}
\end{equation}
The formula for the phase error rate of the single-photon events in $X_A$ is \cite{LCW+14}
\begin{equation}
    \varphi_X:=\frac{c_{X,1}}{y_{X,1}}\leq \frac{v_{Z,1}}{y_{Z,1}}+\gamma\Big(\varepsilon_{sec},\frac{v_{Z,1}}{y_{Z,1}},y_{Z,1},y_{X,1}\Big),
    \label{eq_psix}
\end{equation}
where
\begin{equation}
    \gamma(a,b,c,d):=\sqrt{\frac{(c+d)(1-b)b}{cd\log2}}\log_2\Big(\frac{c+d}{cd(1-b)b}\frac{21^2}{a^2}\Big).
    \label{eq_gamma}
\end{equation}

\subsubsection{Asymptotic analysis of key rate per pass} 
The asymptotic key length is determined by increasing the number of CubeSat passes. Let $M$ denote the total number of CubeSat passes than the asymptotic secret key length is given by $l_\infty = \lim\limits_{M \to \infty} \frac{l_M}{M}$ \cite{SBM+22} where \( l_M \) represents the secret key length (SKL) achieved from \( M \) CubeSat passes. The quantity \( l_\infty \) is determined by analyzing the asymptotic scaling of the ratio \( \frac{l_M}{M} \).
The estimation of vacuum counts per pass is expressed as \cite{SBM+22}:
\begin{equation}
    \frac{y_{X(Z),0}}{M}= \frac{\tau_0}{\mu_2-\mu_3}\Big(\frac{\mu_2\Gamma_3 (n_{X(Z),3}-\mathrm{\delta}^-_{X(Z),3})-\mu_3\Gamma_2 (n_{X(Z),2}+\mathrm{\delta}^+_{X(Z),2})}{M}\Big),
    \label{eq_yx0/m}
\end{equation}
where $n_{X(Z),j}$ represents the number of sifted counts in the $X(Z)$ basis from pulses with intensity $j$. The term $\tau_0$ denotes the average probability that the laser transmits a vacuum state. Additionally, $\Gamma_j = \exp(\mu_j)/p_j$ and $\mathrm{\mathrm{\delta}}_{X(Z),j}^\pm$ are determined using the multiplicative Chernoff bound \cite{YZG+20}. The asymptotic behavior of these correction terms follows the scaling \( O\Big(\sqrt{n_{X(Z)}}\Big) \), which implies that the scaling with respect to the number of CubeSat passes is \( O(\sqrt{M}) \). As a result, the terms \( \frac{\mathrm{\delta}_{X(Z),j}^\pm}{M} \) scale as \( O\Big(\frac{1}{\sqrt{M}}\Big) \), and consequently, they approach zero as \( M \to \infty \).
 As expected, the finite statistical correction term diminish in this limit. Assuming each CubeSat pass follows the same orbit, the total number of counts \( n_{X(Z),j} \) can be expressed as \( M \) times the number of counts for a single pass, \( n^{(1)}_{X(Z),j} \).  from this we obtain \cite{SBM+22}.

\begin{equation}
 \lim\limits_{M \to \infty}\frac{y_{X(Z),0}}{M}= \frac{\tau_0}{\mu_2-\mu_3}\Big(\mu_2\Gamma_3 ( n^{(1)}_{X(Z),3} )-\mu_3\Gamma_2 ( n^{(1)}_{X(Z),2} )\Big)= y^\infty_{X(Z),0} , 
 \label{eq_yx0/m_inf}
\end{equation}
where $y_{X(Z),0}^\infty$ represents the asymptotic estimate of the vacuum counts. For a single transmission pass, which will be formally defined in the next paragraph, the key rate for the efficient BB84 protocol under asymptotic conditions can be determined. By applying a similar methodology to each term in $l_M/M$, the asymptotic key rate can be written as
\begin{equation}
R_\infty^{eff} = \frac{l_\infty }{N_p} =  \left\lfloor y_{X,0}^\infty+y_{X,1}^\infty(1-h(\varphi_X^\infty))-\lambda_{ec}^\infty\right\rfloor.
\label{eq_reff_inf}
\end{equation}
The phase error rate, denoted as \( \varphi_X^\infty \), is given by the ratio \( \frac{v_{Z,1}^\infty}{y_{Z,1}^\infty} \), where \( y_{X,1}^\infty \), \( y_{Z,1}^\infty \), and \( v_{Z,1}^\infty \) represent the asymptotic estimates for the single-photon counts in the \( X \)-basis, the \( Z \)-basis, and the number of single-photon errors in the \( Z \)-basis, respectively, for a single pass.
 These asymptotic quantities, including $v_{Z,1}^\infty$, $y_{X,0}^\infty$, and $y_{X,1}^\infty$, are determined by averaging the single-pass values over an infinite number of passes. A refined estimate of the error correction term, $\lambda_{ec}$, and its asymptotic upper bound is provided in \cite{TMP+17}, from which it follows that $\lambda_{ec}^\infty = n_X^{(1)} h(Q)$, where $Q$ is the QBER for a single pass. Similarly the asymptotic key rate for the standard BB84 protocol can be determined by utilizing both the $X$ and $Z$ bases for key generation and parameter estimation and can be written as-
\begin{equation}
R_\infty^s = \frac{l_\infty}{N_p} = \frac{1}{2} \left\lfloor y_{X,0}^\infty+(y_{X,1}^\infty(1-h(\varphi_X^\infty)))-\lambda_{ecX}^\infty+ y_{Z,0}^\infty+(y_{Z,1}^\infty(1-h(\varphi_Z^\infty)))-\lambda_{ecZ}^\infty\right\rfloor.
\label{eq_rs_inf}
\end{equation}

\subsection{Elliptical Beam Approximation Model in CubeSat-Based Optical Links}
\label{Elliptical Beam Approximation Model in Satellite-Based Optical Links}
\begin{figure}
    \centering
    \includegraphics[width=\linewidth]{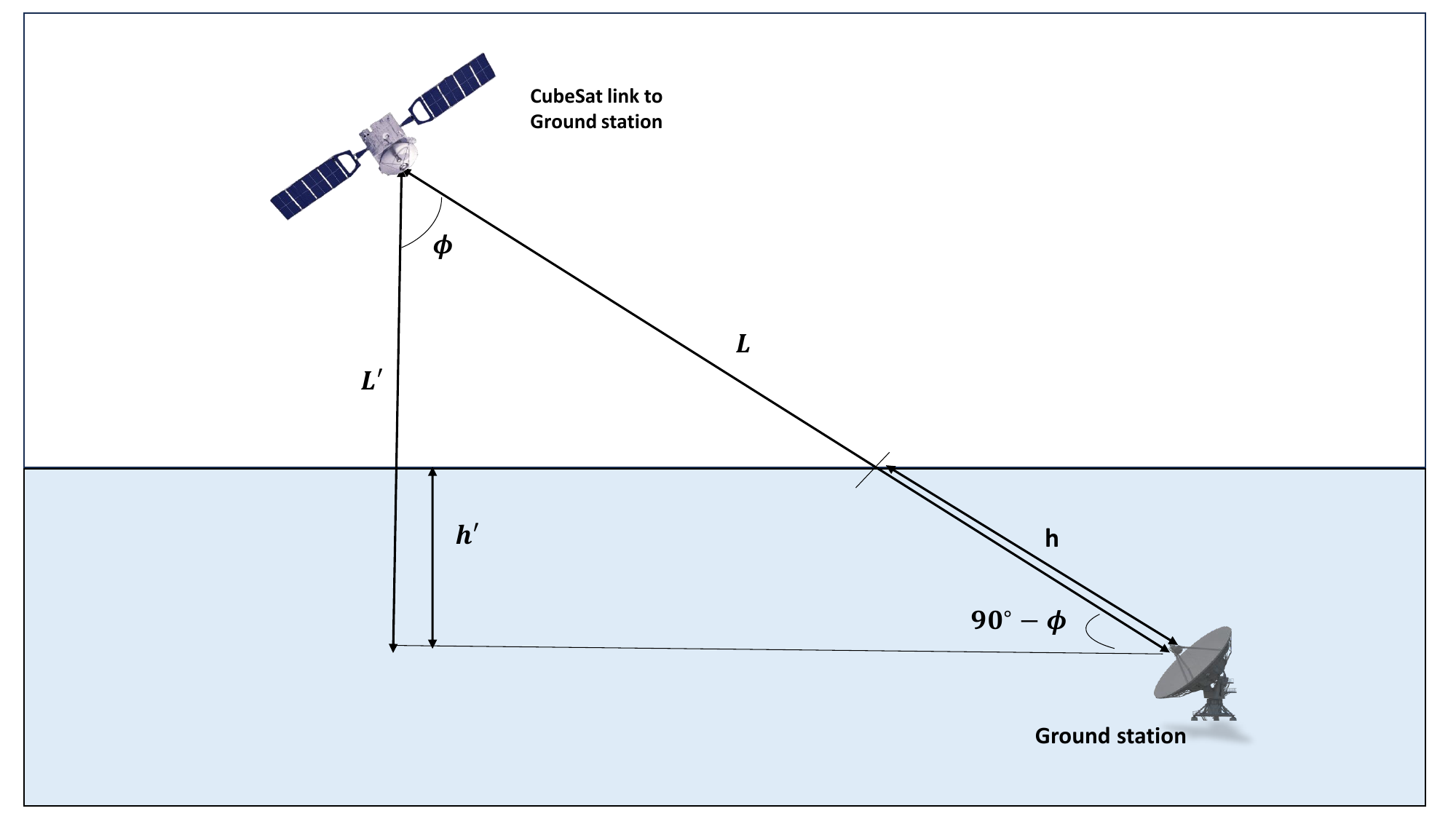}
    \caption{The diagram illustrates the free-space optical link, which is non-uniform, between the ground station and the CubeSat. Key parameters include \( h' \), the atmosphere's thickness, and \( L' \), the CubeSat's altitude. The length of light propagation through the atmosphere is denoted by $h$, while $L$ represents the total link length between the CubeSat and the ground station. The zenith angle is indicated by $\phi$. The downlink configuration describes the transmission of optical signals from the CubeSat to the ground station.}
    \label{cubesat}
\end{figure}

This study focuses on evaluating the key rate performance for both the efficient BB84 and standard BB84 protocols using two decoy states under various weather conditions for CubeSat. To model channel transmission, we employ the elliptical-beam approximation for atmospheric links as introduced by Vasylyev et al \cite{VSV16, VSV+17}. Additionally, we adopt a generalized approach alongside the varying weather conditions presented in \cite{LKB19}. This methodology significantly influences the transmittance values, as transmittance of the channel depends on the characteristics of the beam and the size of the receiving aperture.\\
The atmosphere comprises multiple layers, each characterized by distinct physical properties such as air density, temperature,  pressure and ionized particles. These layers vary in thickness depending on the location. To simplify the analysis, a model is adopted for a satellite-based optical link, where a homogeneous atmosphere is assumed up to a particular altitude, \( h' \), after which a vacuum persists until the CubeSat at altitude \( L' \), as shown in Fig. \ref{cubesat}. Instead of representing physical quantities as continuous altitude functions, this approach emphasizes two key elements: the effective altitude range, \( h' \) and the value of the physical property in the uniform atmosphere. This approach is considered highly reliable, as atmospheric effects are most prominent in the first 10 to 20 kilometers above the ground.
 In our study, we have chosen $L$ to be $400$ $km$ for CubeSats \cite{LVM+23} and assumed that the zenith angle ranges from $[0^\circ,66^\circ]$.\\
Now considering the transmittance, as defined in Eq. (\ref{eq_eeta}), for an elliptical beam that strikes on a circular aperture of radius \( r_a \), it can be represented mathematically as \cite{LKB19}:

 \begin{equation}
     \eta(x_0,y_0,W_1,W_2,\alpha) = \frac{2 \, \chi_{ext}}{\pi W_1 W_2}\int_{0}^{r_a} \rho \, d\rho \int_{0}^{2\pi} d\theta e^{-2A_1(\rho\cos\! \theta-\rho_0)^2}\times e^{-2A_2\rho^2\sin^2\! \theta}\times e^{-2A_3(\rho\cos\! \theta-\rho_0)\rho\! \sin\! \theta}.
     \label{eq_eeta}
 \end{equation}
 {The parameters \( x_0 \) and \( y_0 \) represent the beam centroid coordinates,  {$W_1$ and $W_2$ indicate the semi-principal axes of the beam's elliptical profile, and $\alpha$ represents the angle of orientation of the elliptical beam.} These beam parameters, along with the aperture radius \( r_a \), determine the transmittance.} Here, $\rho$ and $\theta$   signify the polar coordinates of the $\rho$ vector (see details in \nameref{App. A}).

 In the subsequent section, we will analyze the performance of the selected protocols for CubeSat-based links. This analysis requires the computation of average key rates over the probability distribution of the transmittance (PDT) \cite{LKB19}, evaluated for various link lengths and configurations. This process can be expressed as \cite{LKB19}.
\begin{equation}
    R_{avg}= \int_{0}^{1} R(\eta) P(\eta) d\eta = \sum_{i=1}^{N_{bins}} R(\eta_i) P(\eta_i).
    \label{eq_ravg}
\end{equation}
In this context, $R_{avg}$ represents the average key rate, while $R(\eta)$ denotes the key rate for a specific transmittance value. The probability distribution of transmittance (PDT), represented by $P(\eta)$, is employed. To calculate the integral average, the interval $[0,1]$ is divided into $N_{\text{bins}}$ subintervals, each centered at $\eta_i$, where $i$ spans from 1 to $N_{\text{bins}}$. The weighted key rates are summed to obtain the average. The values of $P(\eta_i)$ are obtained through random sampling, as described in the previous section.  The specific expressions for the different key rate implementations, $R(\eta)$, are described in Section~\ref{Finite key rate and Asymptotic analysis} (see Eqs.~\ref{eq_reff}, \ref{eq_rs}, \ref{eq_reff_inf}, and \ref{eq_rs_inf}).

\section{Results and Discussion}
\label{Performance Analysis of QKD Implementation in CubeSat System}
\begin{figure}[htbp!]
    \centering
    \begin{subfigure}[b]{0.48\textwidth}
        \centering
        \includegraphics[width=\textwidth]{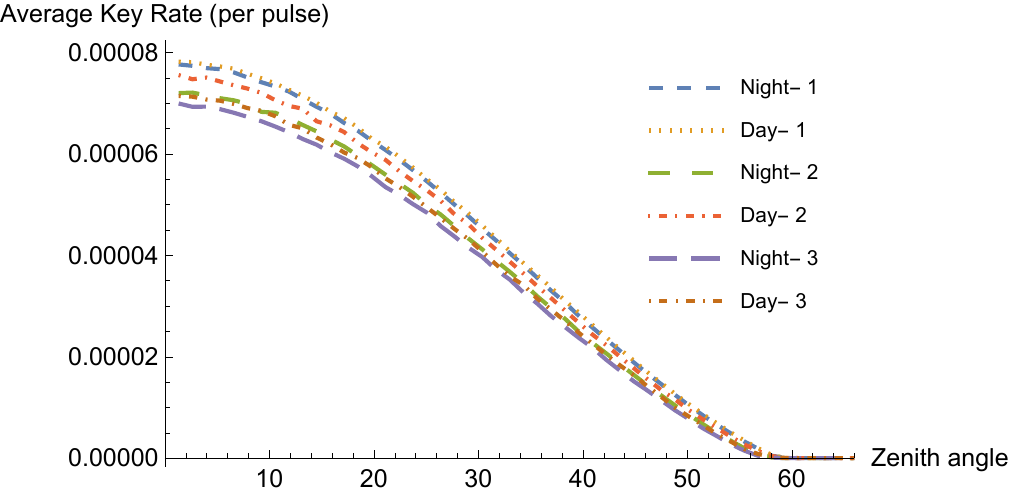} 
         \caption{}
       \label{fe2a}
    \end{subfigure}
    \hfill
    \begin{subfigure}[b]{0.48\textwidth}
        \centering
        \includegraphics[width=\textwidth]{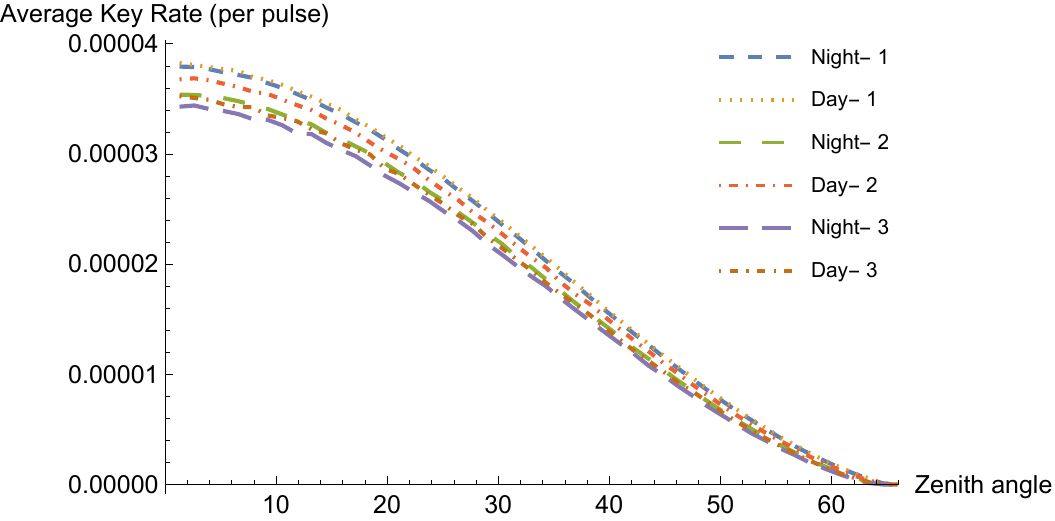} 
        \caption{}
        \label{fs2b}
    \end{subfigure}

    \vskip\baselineskip 

    \begin{subfigure}[b]{0.48\textwidth}
        \centering
        \includegraphics[width=\textwidth]{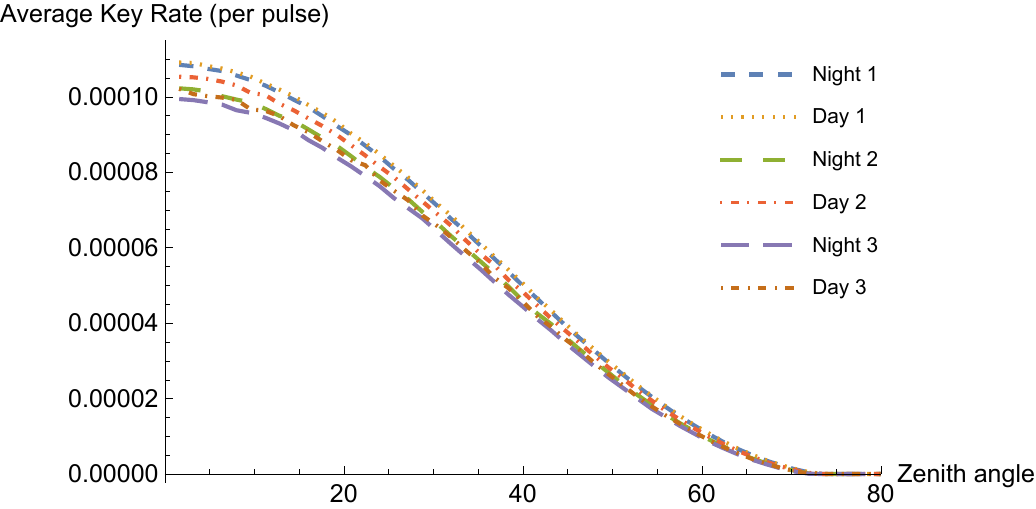} 
        \caption{}
        \label{ae2c}
    \end{subfigure}
    \hfill
    \begin{subfigure}[b]{0.48\textwidth}
        \centering
        \includegraphics[width=\textwidth]{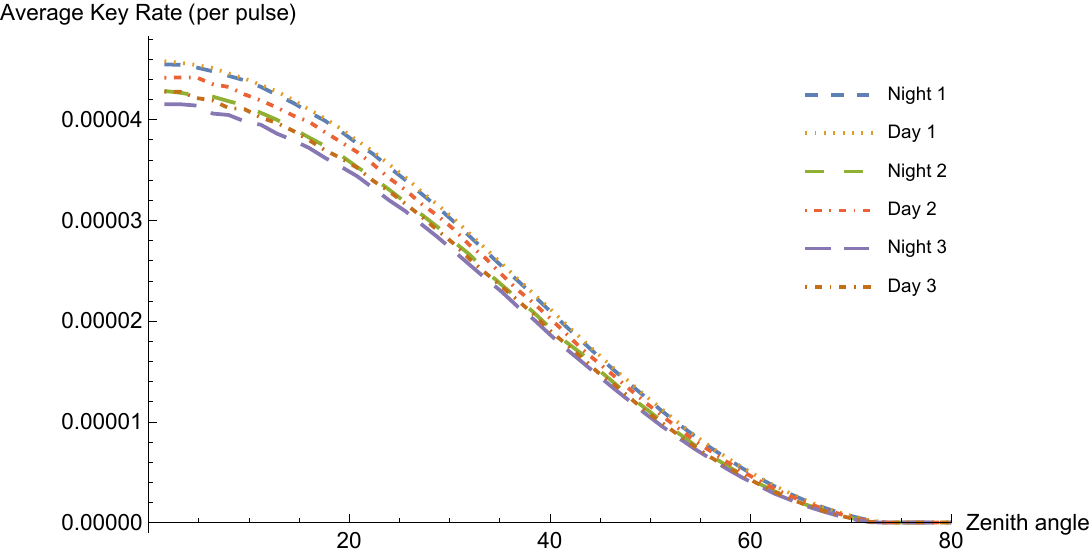} 
        \caption{}
        \label{as2d}
    \end{subfigure}
    \caption{The plot illustrates how the average key rate (per pulse) varies with the zenith angle in a downlink configuration across six distinct weather scenarios: clear (Day 1), moderate wind (Day 2), windy (Day 3), as well as clear (Night 1), slightly foggy (Night 2), and moderately foggy (Night 3). The upper row represents the finite key rate, and the lower row corresponds to asymptotic key rate. Figure 2(a) and 2(b) shows the average finite key rate for the efficient and standard BB84 protocols, respectively, while Figure 2(c) and 2(d) shows the average asymptotic key rate for the efficient and standard BB84 protocols, respectively under the weather conditions (Day 1/2/3 and Night 1/2/3).
}

    \label{fig2}
\end{figure}

 {This section examines the average key rate as a function of the zenith angle using PDT, obtained after performing the weighted sum, for the efficient and standard BB84 protocols in CubeSat-based QKD under different weather conditions in a downlink scenario. To accurately model atmospheric losses, we employ the elliptical beam approximation, which effectively captures the impact of beam spreading and turbulence-induced distortions on photon transmission. Additionally, we analyze the PDR for different zenith angles, considering both protocols in the finite and asymptotic cases.} CubeSats typically feature compact optics with apertures of $\leq 10\,\text{cm}$ \cite{LKB19}. We present the results of numerical simulations for CubeSat-based implementations of the efficient and standard BB84 protocols, evaluated under finite key and asymptotic key analyses. The simulations incorporate the experimental parameters specified in table \ref{table} \cite{SBM+22, LKB19}. These analyses consider varying atmospheric conditions, including clear, slightly foggy, and moderately foggy nights, as well as non-windy, moderately windy, and windy days~\cite{LKB19}. In this scenario, the critical factors include both atmospheric effects, the transmitter and receiver telescope radii, and the signal wavelength for CubeSats in orbit. For the CubeSat, a radius of $r_{\text{sat}} = 5\,\text{cm} \, (W_0)$ is considered, whereas the ground station telescope has a radius of $r_{\text{grnd}} = 50\,\text{cm}$, and the signal wavelength is $\lambda = 785\,\text{nm}$. We have opted for a downlink configuration due to its lower transmission losses \cite{DMB+24}.\\
\begin{table}[h!]
\centering
\caption{Parameters related to the baseline SatQKD system, various atmospheric weather conditions, and the optical and technical characteristics of the link.}
\begin{tabularx}{\textwidth}{>{\raggedright\arraybackslash}X >{\raggedright\arraybackslash}X >{\raggedright\arraybackslash}X}
\hline
\textbf{Parameter} & \textbf{Value} & \textbf{Short Description} \\
\hline
$\mathcal{W}_0$ & 5 cm & CubeSats Down-link \\
$r_a$ & 50 cm & CubeSats Down-link \\
$\lambda$ & 785 nm & Wavelength of the signal light \\
$\beta$ & 0.7 & Parameter in $\chi_{\text{ext}}(\phi)$ \\
$p_e$ & $2 \times 10^{-6}$ rad & Pointing error \\
$h'$ & $20$ km & Atmosphere thickness \\
$L'$ & $400$ km & Minimum altitude (at zenith) \\
$n_0$ & 0.61 m$^{-3}$ & Night-~1 \\
$n_0$ & 0.01 m$^{-3}$ & Day-~1 \\
$n_0$ & 3.00 m$^{-3}$ & Night-~2 \\
$n_0$ & 0.05 m$^{-3}$ & Day-~2 \\
$n_0$ & 6.10 m$^{-3}$ & Night-~3 \\
$n_0$ & 0.10 m$^{-3}$ & Day-~3 \\
$C_n^2$ & $1.12 \times 10^{-16}$ m$^{-2/3}$ & Night-~1 \\
$C_n^2$ & $1.64 \times 10^{-16}$ m$^{-2/3}$ & Day-~1 \\
$C_n^2$ & $5.50 \times 10^{-16}$ m$^{-2/3}$ & Night-~2 \\
$C_n^2$ & $8.00 \times 10^{-16}$ m$^{-2/3}$ & Day-~2 \\
$C_n^2$ & $1.10 \times 10^{-15}$ m$^{-2/3}$ & Night-~3 \\
$C_n^2$ & $1.60 \times 10^{-15}$ m$^{-2/3}$ & Day-~3 \\
$QBER_I$ & $5 \times 10^{-3}$ & Intrinsic QBER \\
$p_{ap}$ & $1 \times 10^{-3}$ & Afterpulse probability \\
$p_{ec}$ & $5 \times 10^{-7}$ & Extraneous count probability/pulse \\
$f_s$ & $1 \times 10^{8}$  {Hz} & Source rate \\
$\varepsilon_{corr}$ & $ 10^{-15}$ & Correctness parameter \\
$\varepsilon_{sec}$ & $ 10^{-9}$ & Secrecy parameter \\
$\lambda_{ec}$ &  depends on block size (see text) &  Error correction efficiency   \\
\hline
\end{tabularx}
\label{table}
\end{table}
The Eqs. (\ref{eq_reff}) and (\ref{eq_rs}) represent the finite key rate expressions, while Eqs. (\ref{eq_reff_inf}) and (\ref{eq_rs_inf}) provide the asymptotic key rate formulations for the efficient and standard BB84 protocols respectively. By incorporating the PDT in CubeSat-based communication, these expressions enable the computation of the average key rate for CubeSat-based quantum communication systems. Here, Fig. \ref{fig2} depicts the dependence of the average key rate on the zenith angle, incorporating the PDT. The analysis is performed for a downlink scenario across various weather conditions, as outlined in table \ref{table}. Each value on the plot is determined from 1,000 samples of the parameters, based on Eq. (\ref{eq_vpara}) in \nameref{App. A} and computed using Eq. (\ref{eq_eeta}). {Figure. \ref{fig2} represents the finite and asymptotic key analysis, demonstrating that the secure key rate for efficient BB84 is generally higher than that of standard BB84, particularly during daytime conditions compared to nighttime.
 In Fig. \ref{fe2a}, at the zenith position (i.e., \(0^\circ\) zenith angle), the efficient BB84 protocol achieves a key rate of approximately \(8 \times 10^{-5}\) per pulse under clear daytime conditions (Day 1), whereas under clear nighttime conditions (Night 1), the key rate slightly decreases to \(7.7 \times 10^{-5}\) per pulse. This reduction is primarily due to increased aerosol scattering and the potential formation of haze at night due to low temperature. In Fig. \ref{fs2b}, the standard BB84 protocol follows a similar trend, with key rates of approximately \(4 \times 10^{-5}\) per pulse in Day 1 and \(3.8 \times 10^{-5}\) per pulse in Night 1, yielding approximately half the key rate of the efficient BB84 protocol in both cases. The disparity in key rates becomes more pronounced at higher zenith angles, such as \(50^\circ\). At a zenith angle of \(50^\circ\), the reduction in key rate compared to the zenith position is significant for both the efficient and standard BB84 protocols. This decline can be theoretically attributed to the increased path length at larger zenith angles. As the zenith angle increases, the signal traverses a longer atmospheric path, leading to greater attenuation due to absorption and scattering. Furthermore, enhanced beam divergence reduces the overlap with the receiving telescope aperture, further lowering the detected photon count. In Fig. \ref{fe2a}, for the efficient BB84 protocol, the key rate decreases from \(8 \times 10^{-5}\) per pulse at zenith position to \(1.09 \times 10^{-5}\) per pulse at \(50^\circ\) during Day 1, representing approximately a 7.34-fold reduction. Similarly, during Night 1, the key rate decreases from \(7.7 \times 10^{-5}\) per pulse at zenith position to \(1.06 \times 10^{-5}\) per pulse at \(50^\circ\), yielding a 7.26-fold reduction. In Fig. \ref{fs2b}, for the standard BB84 protocol, the key rate drops from \(4 \times 10^{-5}\) per pulse at \(0^\circ\) to \(7.8 \times 10^{-6}\) per pulse at \(50^\circ\) during Day 1, corresponding to a 5.13-fold reduction. In Night 1, the key rate decreases from \(3.8 \times 10^{-5}\) per pulse at zenith position to \(7.6 \times 10^{-6}\) per pulse at \(50^\circ\), leading to a 5-fold reduction. These reductions highlight the significant impact of increasing zenith angle on the key rate, primarily due to enhanced atmospheric losses and longer path lengths through the atmosphere at higher zenith angles. Furthermore, the results consistently demonstrate the superior performance of the efficient BB84 protocol over the standard BB84 protocol, as it achieves significantly higher key rates under both conditions.}  Across all the plots in Fig. \ref{fig2}, the pattern of the plots, corresponding to the sequential arrangement of different weather conditions, remains consistent for both protocols.  The sequence of weather conditions yielding higher key rate values follows the order: day-~1, night-~1, day-~2, night-~2, day-~3, and night-~3. A key aspect of interest is the comparison of system performance between daytime and nighttime operations.
 
\begin{figure}[htbp!]
    \centering
    \begin{subfigure}[b]{0.48\textwidth}
        \centering
        \includegraphics[width=\textwidth]{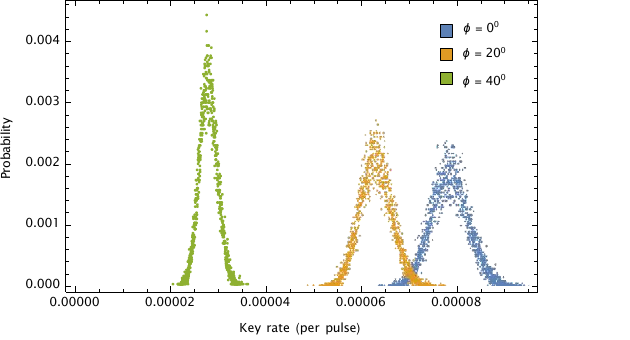} 
         \caption{}
       \label{pdrfe3a}
    \end{subfigure}
    \hfill
    \begin{subfigure}[b]{0.48\textwidth}
        \centering
        \includegraphics[width=\textwidth]{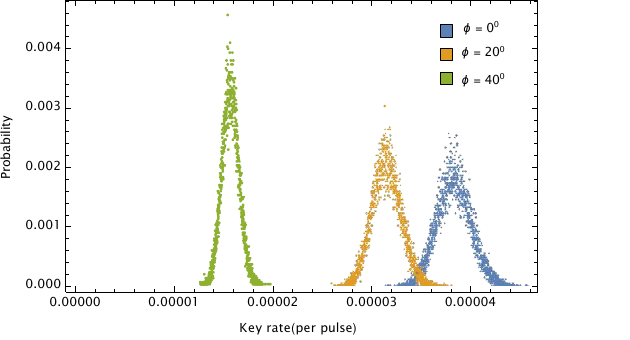} 
        \caption{}
         \label{pdrfs3b}
    \end{subfigure}

    \vskip\baselineskip 

    \begin{subfigure}[b]{0.48\textwidth}
        \centering
        \includegraphics[width=\textwidth]{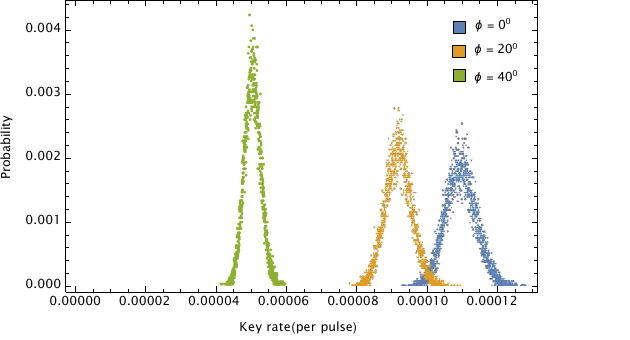} 
        \caption{}
        \label{pdrae3c}
    \end{subfigure}
    \hfill
    \begin{subfigure}[b]{0.48\textwidth}
        \centering
        \includegraphics[width=\textwidth]{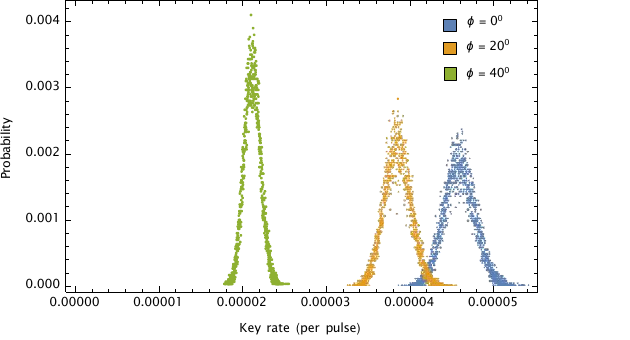} 
        \caption{}
        \label{pdras3d}
    \end{subfigure}

    \caption{Plot depicting the variation in key rate distribution across different zenith angles $(\phi)$ under condition Day-1 a) PDR for Finite efficient BB84 protocol, b) PDR for Finite standard BB84 protocol, c) PDR for Asymptotic efficient BB84 protocol, d) PDR for Asymptotic standard BB84 protocol.}
    \label{pdr3}
\end{figure}
 
In the daytime, elevated temperatures lead to more intense winds and enhanced mixing among various atmospheric layers, leading to more significant turbulence effects. However, on average, clear daytime conditions exhibit less moisture in the lower part of the atmosphere than nighttime conditions which results in reduced beam spreading due to scattering particles. In contrast, cooler nighttime temperatures lead to a less turbulent atmosphere, coupled with formation of haze and mist. As a result, scattering has a more pronounced impact at night than turbulence does during the day.  {The secure key rate exhibits a progressive decline with increasing atmospheric turbulence and fog. Under moderate wind conditions (Day 2) and slightly foggy conditions (Night 2), the efficient BB84 protocol achieves a key rate of \(7.5 \times 10^{-5}\) per pulse during Day 2 and \(7.2 \times 10^{-5}\) per pulse during Night 2 at the zenith position. At a zenith angle of \(50^\circ\), the key rate decreases to the order of \(10^{-6}\), reflecting the impact of increased optical path length and atmospheric attenuation. The standard BB84 protocol follows a similar trend but exhibits a lower key rate, reducing to \(3.6 \times 10^{-5}\) per pulse during Day 2 and \(3.5 \times 10^{-5}\) per pulse during Night 2 at the zenith position. At \(50^\circ\) zenith angle, the standard BB84 protocol experiences a more pronounced decline, highlighting its greater susceptibility to turbulence-induced fluctuations and channel losses compared to the efficient BB84 protocol.  With further intensification of atmospheric disturbances in windy (Day 3) and moderately foggy (Night 3) conditions, the degradation in key rate becomes more severe. The efficient BB84 protocol remains operational across all conditions, sustaining key generation even at higher zenith angles, whereas the standard BB84 protocol approaches the threshold beyond which key generation becomes impractical. The increased optical path length at larger zenith angles exacerbates scattering and absorption effects, leading to heightened attenuation. This degradation is particularly detrimental to the standard BB84 protocol, which exhibits greater sensitivity to statistical fluctuations and environmental losses.  These results reinforce that clear daytime conditions (Day 1) are optimal for CubeSat-based QKD due to minimal scattering losses, while nighttime conditions (Night 1) introduce slightly higher losses due to increased aerosol scattering. However, severe turbulence and fog (Day 3 and Night 3) significantly impair performance, particularly for the standard BB84 protocol. The superior performance of the efficient BB84 protocol aligns with theoretical expectations, as its biased basis selection enhances the sifting ratio, thereby improving parameter estimation accuracy and increasing overall key generation efficiency.} Additionally, the efficient BB84 protocol uses one basis for key generation and the other for parameter estimation, thereby maximizing the utilization of measurement outcomes and minimizing unused data. In contrast, the standard BB84 protocol employs both bases with equal probability and utilizes both for key generation, necessitating parameter estimation for each basis. To estimate the signal parameters, the protocol reveals a random sample of results from each measurement basis separately. Consequently, only half of the revealed results from each basis are used for parameter estimation, introducing greater statistical uncertainty compared to the efficient BB84 protocol.  {In Fig.~\ref{ae2c} and Fig.~\ref{as2d}, the asymptotic regime follows a similar trend to the finite-key regime, with key rates increasing by approximately 1.2 to 1.5 times. However, the relative advantage of the efficient BB84 protocol remains consistent across all conditions, resulting in a higher asymptotic average key rate compared to the standard BB84 protocol. This trend is evident in Fig.~\ref{ae2c} and Fig.~\ref{as2d}, where the asymptotic key rates exceed those in the finite-key regime, as shown in Fig.~\ref{fe2a} and Fig.~\ref{fs2b}, further reinforcing the performance superiority of the efficient BB84 protocol.}  Notably, in the finite-key scenario, the key rate drops to zero earlier at a zenith angle of $62^\circ$, whereas in the asymptotic case, it extends up to $75.2^\circ$. This difference arises because the asymptotic analysis assumes an infinite number of satellite passes, thereby eliminating statistical fluctuations and finite-size effects. As a result, the key rate remains higher and persists to a greater zenith angle compared to the finite-key case.\\
 {In Fig. \ref{pdr3}, we present the PDR at different zenith positions under downlink configuration. In this scenario, we consider the optimal performance under daytime condition 1. A dataset comprising $3 \times 10^4$ beam parameters served to simulate the average key rate, with outcomes rounded to five decimal places for generating the PDR graphs for efficient and standard BB84 protocols in both finite and asymptotic cases. A comparison between $\phi=0^\circ$ and $\phi=40^\circ$ shows that a higher key rate is observed at $\phi=0^\circ$, whereas the maximum probability of the key rate is higher at $\phi=40^\circ$. Notably, a larger key rate correlates with a decreased probability of occurrence. At lower zenith angles (e.g., $\phi = 0^\circ$), the key rate distribution is broader, indicating greater variation in key rates. Shorter optical path lengths result in stronger signal transmission, leading to higher key rates. However, atmospheric effects such as turbulence still introduce fluctuations, causing a wider distribution and reducing the probability of key rate at lower zenith angles. At higher zenith angles (e.g., $\phi = 20^\circ$ and $\phi = 40^\circ$), the probability of the key rate increases, but the key rate distribution becomes narrower, and the overall key rate decreases. The longer atmospheric path length results in higher attenuation, beam spreading, and turbulence effects, reducing transmittance and increasing photon losses. Consequently, the probability distribution of the key rate becomes more concentrated at lower values, reflecting a reduced spread of key rates. The overall key rate declines due to increased channel losses and lower detection efficiency.}\\
 {In both finite and asymptotic cases, the efficient BB84 protocol consistently achieves a higher key rate than the standard protocol. This is evident in the PDR, where the efficient BB84 protocol exhibits a broader distribution at lower zenith angles and maintains higher key rates compared to the standard BB84 protocol. Additionally, as the zenith angle increases, the PDR of the efficient BB84 protocol narrows more gradually compared to the standard BB84 protocol, which experiences a sharper reduction in key rate probability. This trend is consistently observed across all subplots in Fig.~\ref{pdr3}, highlighting the superior performance of the efficient BB84 protocol in CubeSat-based QKD implementations.} 
To highlight the advantages of the efficient BB84 protocol over the standard BB84 protocol, we provide a comparative functional table. Table \ref{table3} outlines key performance metrics, emphasising improvements in key rate, error resilience, and protocol efficiency  \cite{SBM+22}. 
\FloatBarrier
\begin{table*}[h]
    \centering
    \caption{Functional Comparison Between Standard and Efficient BB84 Protocols}
    \renewcommand{\arraystretch}{1.3} 
     \begin{tabular}{|p{5cm}|p{6cm}|p{6cm}|}
        \hline
        \textbf{{Feature}} & \textbf{Standard BB84 Protocol} & \textbf{Efficient BB84 Protocol} \\
        \hline
       \textbf{Basis Selection and Sifting Ratio} & Choose X and Z basis with symmetric probability. & Asymmetric basis choice and enhanced sifting ratio.\\
        \hline
        \textbf{Key Generation and Parameter Estimation} & Uses both bases for key generation and parameter estimation. & One basis for key generation and the other for parameter estimation. \\
        \hline
       \textbf{Classical Communication}  & Requires more classical communication due to separate QBER estimation.  & Requires less classical communication as QBER estimation is implicitly included in basis choice and revealed during sifting. \\
        \hline
        \textbf{Finite Key rate} & Lower average finite key rate across different weather conditions due to less sifting ratio.  & Higher average finite key rate across different weather conditions due to better sifting ratio, resulting in longer raw key length.\\
        \hline
        \textbf{Asymptotic Key rate} & Lower average asymptotic key rate across different weather conditions. & Higher asymptotic key rate across different weather conditions due to enhanced protocol efficiency and better resilience to channel losses.\\
        \hline
        \textbf{Finite Statistical Uncertainty} & Requires parameter estimation for two bases, increasing statistical uncertainty. & Eliminates the need for two-basis parameter estimation, reducing statistical uncertainty and improving key rate estimation.  \\  
        \hline
        \textbf{Probability Distribution of Key Rate} & PDR exhibits a less broader distribution at lower zenith angles, narrowing moderately at higher angles, with lower key rates than efficient BB84 under daytime condition 1. & PDR exhibits a broader distribution at lower zenith angles, gradually narrowing at higher angles while maintaining higher key rates under daytime condition 1. \\  
        \hline
    \end{tabular}
    \label{table3}
\end{table*}
\FloatBarrier

\section{Conclusion}
\label{Conclusion}
{
In this work, we investigated the performance of two quantum key distribution protocols—the efficient BB84 and the standard BB84—under a two-decoy state setting in a CubeSat-based free-space communication downlink scenario. A key aspect of this study is the application of the elliptical beam approximation to model transmittance more accurately compared to the conventional circular beam approach. This refinement provides a more realistic characterization of beam divergence and atmospheric effects, thereby improving the reliability of key rate estimation for CubeSat-based QKD.  

A comprehensive finite and asymptotic key rate analysis has been conducted to evaluate and compare the performance of the efficient and standard BB84 protocols under varying atmospheric conditions in a CubeSat-based QKD system. The analysis utilized the elliptical beam approximation model to accurately characterize channel transmittance and its impact on key rate generation. Our results demonstrate that the efficient BB84 protocol consistently achieves higher key rates across different transmission conditions, particularly in low-transmittance regimes, making it a more suitable candidate for CubeSat-based quantum communication. The dependence of the average key rate per pulse on the ``Zenith angle'' under different atmospheric conditions has been analyzed, along with the PDR. The results indicate that the PDR exhibits a consistent shape across all analyzed scenarios, revealing critical insights into the performance variations of the protocols.  

While the present study utilizes uniform and normal distributions for beam parameter modeling, other transmittance models—such as log-normal, Gamma-Gamma, and Double Weibull distributions—may further enhance accuracy under varying atmospheric conditions. The selection of an appropriate distribution is influenced by factors like the  intensity of turbulence, link distance, and the optical system configuration. Future work will focus on incorporating these alternative models and optimizing finite key generation techniques for higher-dimensional QKD protocols in CubeSat-based quantum communication. Additionally, real-time adaptive beam correction methods could be explored to mitigate the impact of atmospheric fluctuations.  

This study integrates finite and asymptotic key analysis with elliptical beam modeling across various atmospheric conditions, along with PDR, to establish a more precise analytical framework for key rate estimation in CubeSat-based QKD. By enhancing the understanding of CubeSat-based quantum key distribution, it contributes to the advancement of satellite-based quantum communication networks. The findings presented provide a practical foundation for bridging the gap between theoretical modeling and real-world implementation, supporting the future development of secure satellite-based quantum communication systems.}

\color{black} 

\subsection*{Acknowledgements}
Muskan would like to thank CSIR for the fellowship support. Muskan also acknowledged Ramniwas Meena for valuable discussions.
\subsection*{Conflict of Interest}
The authors declare no conflict of interest.
\subsection*{Author Contributions}
S.B. conceptualized the problem. M. performed the majority of the calculations. M. and A.D. verified the results. S.B., M., and A.D.  analyzed the results. All authors contributed to the drafting and final approval of the manuscript.
\bibliography{source}
\bibliographystyle{unsrt}

\appendix

\section*{Appendix A}\label{App. A}
\subsection*{Elliptical Beam Approximation Model}
\begin{figure}
    \centering
    \includegraphics[width=\linewidth]{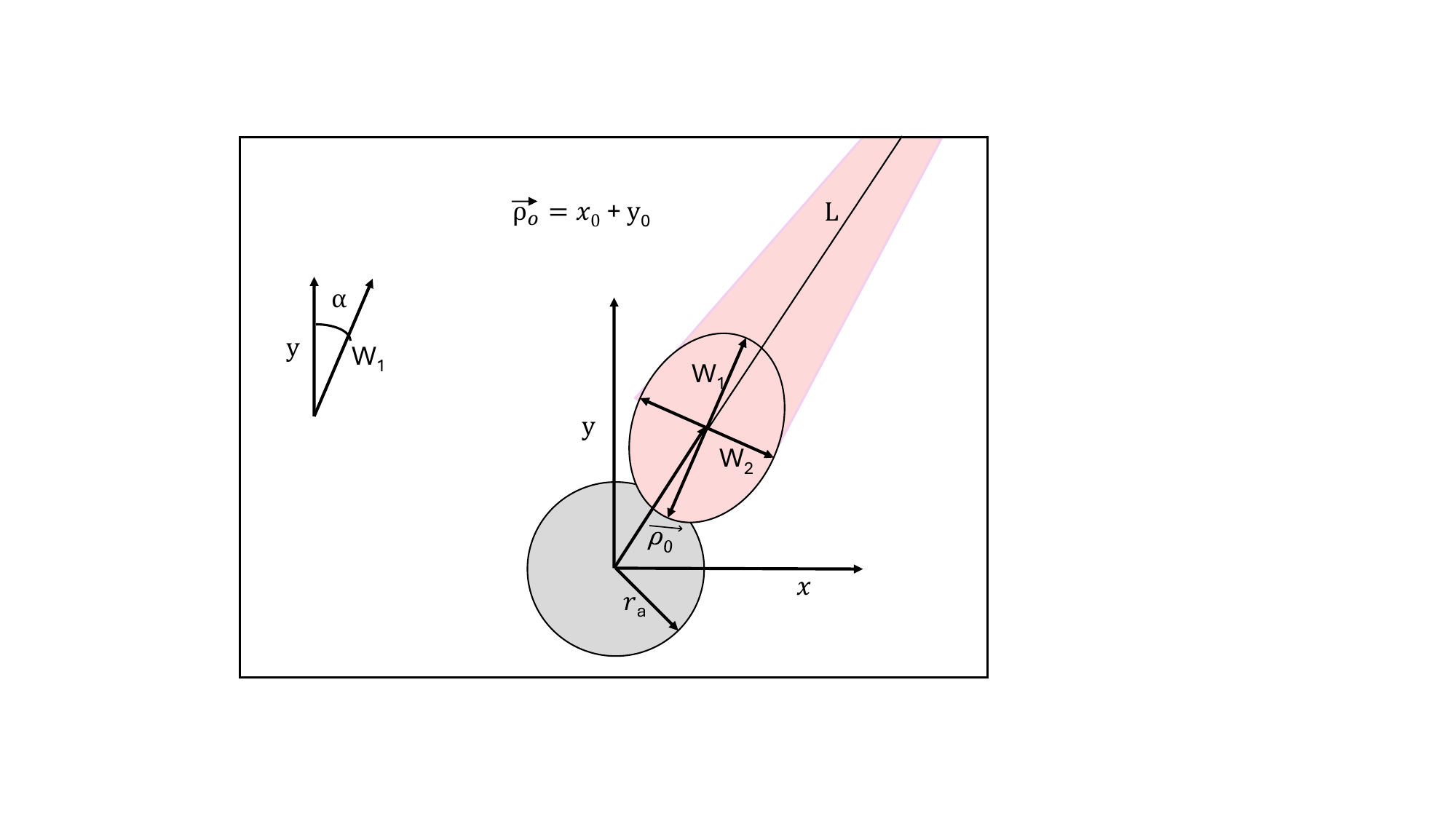}
    \caption{A diagram depicting the receiving aperture and the corresponding received beam. The total propagation link length is denoted as $L$, and $r_a$ represents the radius of the receiving aperture. The beam centroid is positioned at $\rho_0 = (x_0, y_0)$. The elliptical beam profile is defined by its principal semi-axes, $W_1$ and $W_2$, and its orientation angle $\alpha$.}
    \label{elliptical}
\end{figure}

Fluctuations in temperature and pressure over time and space within turbulent atmospheric flows cause stochastic changes in the refractive index of air, leading to transmission losses in photons detected by a receiver with a limited aperture. This turbulence degrades the transmitted signal through effects such as beam wandering, broadening, and deformation. For analysis, we consider a ``Gaussian beam'' that propagates along the $z$-axis and encounters an aperture plane located at $z = L$.
 Assuming perfect ``Gaussian'' beams from the transmitter is idealized; in practice, standard telescopes produce intensity distributions that approximate a circular Gaussian profile with edge truncation effects. These imperfections contribute to beam broadening due to diffraction, which is mitigated by adjusting the initial beam width parameter $(W_0)$ to capture increased far-field divergence. The model includes elliptical beam transmission through a circular aperture and considers the statistical characteristics of this beam as it propagates through turbulence, using a Gaussian approximation. Notably, we assume isotropic atmospheric turbulence for simplicity. For further detail, readers can refer to the Supporting Information in \cite{VSV16}. The quasi-Gaussian beam traverses a link across both atmospheric and vacuum segments, from an orbiting transmitter or a ground station, where the link conditions vary. The received intensity transmittance through a circular aperture of radius $r_a$ in the receiving telescope is generally given as follows \cite{VSV16,DBD+24}:
\begin{equation}
\eta = \int_{|\rho|^2 = r_a^2} d^2 \rho \, |u(\rho, L)|^2.
\label{eq_eeta_rho}
\end{equation}
The term $u(\rho,L)$ represents the beam envelope at the receiver plane, located at a distance $L$ from the transmitter, while $|u(\rho,L)|^2$ denotes the normalized intensity across the entire $\rho-plane$. In the transverse plane, \( \rho \) represents the position vector, while the beam's configuration at the receiver plane is fully defined by the vector parameter \( v \) (see Fig. \ref{elliptical}).

\begin{equation}
    v=(x_0,y_0,W_1,W_2,\alpha).
    \label{eq_vpara}
\end{equation}
The discussion is further extended under the assumption that the parameters characterizing atmospheric effects remain constant with values exceeding zero inside the atmosphere and are assigned zero outside of it. In this context, we can also rely on the assumption that in down-link case \cite{LKB19},
\begin{equation}
\begin{split}
  C_n^2(z) &= C_n^2 \Omega(z - (L - h)) \\
  n_0(z) &= n_0 \Omega(z - (L - h)),
\end{split}
\label{eq_cn_n0}
\end{equation}
where \( C_n^2 \) denotes the refractive index structure constant~\cite{V80} and \( n_0 \) represents the density of scattering particles~\cite{TP88,T84}. The Heaviside step function is represented by $\Omega(z)$ \footnote{This function belongs to the broader class of step functions and takes a value of zero for negative values and one for positive values.}
 As illustrated in Fig. \ref{cubesat}, the parameter \( z \) represents the ``longitudinal coordinate'', \( L \) denotes the total link length, and \( h \) specifies the propagation distance inside the atmosphere.\\
The parameters used in Eq. \ref{eq_eeta} are described as:

\begin{equation}
\begin{split}
     x = \rho\cos\! \theta \\
     y = \rho\sin\! \theta
\end{split}
\label{eq_xy}
\end{equation}
Here, $\rho_0$ and $\theta_0$ denote the polar coordinates associated with the vector $\rho_0$.
\begin{equation}
\begin{split}
     x_0 = \rho_0\cos\! \theta_0 \\
     y_0 = \rho_0\sin\! \theta_0
\end{split}
\label{eq_x0_y0}
\end{equation}
and
\begin{equation}
\begin{aligned}
   A_1 &= \Big(\frac{\cos^2(\alpha-\theta_0)}{W_1^2} + \frac{\sin^2(\alpha-\theta_0)}{W_2^2}\Big) \\
   A_2 &= \Big(\frac{\sin^2(\alpha-\theta_0)}{W_1^2} + \frac{\cos^2(\alpha-\theta_0)}{W_2^2}\Big) \\
   \hspace{-1cm} A_3 &= \Big(\frac{1}{W_1^2} - \frac{1}{W_2^2}\Big) \sin 2(\alpha-\theta_0)
\end{aligned}
\label{eq_a1_a2}
\end{equation}
These expressions are applicable for computational integration, as illustrated in Eq. (\ref{eq_eeta}), utilizing the ``Monte Carlo'' approach or other appropriate techniques. To facilitate integration through the Monte Carlo method, $N$ sets of values for the vector $v$ (refer to Eq. (\ref{eq_vpara}) need to be generated. It is assumed that the angle $(\alpha - \theta_0)$ is uniformly distributed over the interval $[0, \frac{\pi}{2}]$, while the other parameters \footnote{To evaluate transmittance, $W_i$ is first determined from $\Theta_i$ using the expression $\Theta_i = \ln\left(\frac{W_i^2}{W_0^2}\right)$, where $i = 1,2$. The parameter $W_0$ denotes the beam spot radius at the transmitter.}$(x_0, y_0, \Theta_1, \Theta_2)$  follow a normal distribution \cite{WHW+18}. By substituting the simulated values of $v$ into Eq. (\ref{eq_eeta}), numerical integration can be performed. The result of this process includes the extinction factor \cite{LKB19}, $\chi_{\text{ext}}$, leading to $N$ values of atmospheric transmittance, denoted as $\eta(v_i)$, where $i$ ranges from $1$ to $N$.
\end{document}